\documentclass[prd,aps,showpacs,nofootinbib]{revtex4}
\usepackage{amssymb}
\usepackage{graphicx}
\usepackage{amsmath}

\newcommand{\be}{\begin{equation}}
\newcommand{\ee}{\end{equation}}
\newcommand{\bq}{\begin{eqnarray}}
\newcommand{\eq}{\end{eqnarray}}

\begin{document}

\title{Testing Lorentz symmetry violation with an invariant minimum speed} 
\author{Cl\'audio Nassif, A. C. Amaro de Faria Jr. and Rodrigo Francisco dos Santos\\
claudionassif@yahoo.com.br}
\altaffiliation{{\bf CPFT}: Centro de Pesquisas em F\'isica Te\'orica: Rua Rio de Janeiro 1186/s.1304, Lourdes, Belo Horizonte-MG,
 CEP:30.160-041, Brazil.\\
 {\bf IEAv}: Instituto de Estudos Avan\c{c}ados: Rodovia dos Tamoios Km 099, S\~ao Jos\'e dos Campos-SP, CEP:12.220-000, Brazil.\\ 
 {\bf UFF}: Universidade Federal Fluminense: Av. Litoranea s/n, Gragoat\'a, CEP:24210-340, Niter\'oi-RJ, Brazil.}

\date{\today}

\begin{abstract}
This work presents an experimental test of Lorentz invariance violation in the infrared (IR) regime by means of an invariant minimum 
speed in the spacetime and its effects on the time when an atomic clock given by a certain radioactive single-atom (e.g.: isotope $Na^{25}$) is a thermometer for a ultracold
gas like the dipolar gas $Na^{23}K^{40}$. So, according to a Deformed Special Relativity (DSR) so-called Symmetrical Special 
Relativity (SSR), where there emerges an invariant minimum speed $V$ in the subatomic world, one expects that the proper time of such a clock moving close to $V$ in thermal equilibrium with the
ultracold gas is dilated with respect to the improper time given in lab, i.e., the proper time at ultracold systems elapses faster than
the improper one for an observer in lab, thus leading to the so-called {\it proper time dilation} so that the atomic decay rate of a ultracold 
radioactive sample (e.g: $Na^{25}$) becomes larger than the decay rate of the same sample at room temperature. This means a suppression of
the half-life time of a radioactive sample thermalized with a ultracold cloud of dipolar gas to be investigated by NASA in the Cold 
Atom Lab (CAL).
\end{abstract}

\maketitle

\section{\label{sec:level1} Introduction}

In the last 20 years the scientific community has shown an intense interest in the theories that contained and investigated the breakdown
of Lorentz symmetry in many scenarios\cite{Bluhm}\cite{Carroll}\cite{Kostelecky}\cite{Kostelecky1}\cite{Jacobson}\cite{Jun}
\cite{Almeida}\cite{Schreck}\cite{Tiberio}\cite{Belich}\cite{Perennes}\cite{Lambiase} and also the so-called Deformed 
Special Relativities (DSR)\cite{Smolin}\cite{Camelia}, although no relevant experimental fact has demonstrated Lorentz symmetry  
breaking until the present time. However, there could be the evidence that the breakdown of Lorentz symmetry may exist in 
a lower energy regime. In this paper we propose an experimental investigation of this fact in view of a possible existence of a non-null minimum speed $V$ that has been postulated 
in a previous work\cite{N2016} and incorporated into the so-called Symmetrical Special Relativity (SSR) theory, where there 
should be two invariant speeds, namely the speed of light $c$ (the maximum speed) for higher energies, being related to a maximum 
temperature (Planck temperature $T_P\sim 10^{32}K$) within the cosmological scenario (early universe), and the minimum speed 
$V(\sim 10^{-14}m/s)$\cite{N2016} for lower energies, being related to a minimum temperature. However, due to the
unattainable minimum speed $V$ for representing the zero energy of a particle, the new dynamics in SSR subtly changes the conception
about what should mean {\it a minimum temperature} in the cosmological scenario, which will be denominated as the 
{\it Planck scale of temperature} in SSR, i.e., $T_{min}<T<T_P$ as it will be investigated later.  

The theory of SSR is one of the most recent DSRs and it has many far-reaching consequences. As for instance, the Heisenberg uncertainty 
principle that acquires an origin from the spacetime of SSR\cite{N2012} and the tiny positive value of the cosmological constant connected 
to the low vacuum energy density were elucidated by SSR\cite{N2016}\cite{N2015}\cite{N2012}. Also, an intriguing consequence of this minimum speed is the
symmetry between the proper and improper time, where both can either suffer contraction or dilation. This issue is investigated in the present paper
with the propose of establishing an experimental route of detecting a new time effect when the speed $v$ is close to the minimum 
speed $V$ connected to a minimum temperature\cite{N2010}, by taking into account a ultracold gas in thermal equilibrium with a 
sample of radioactive atoms. 

In fact the striking implication concerning the existence of the minimum speed $V$ is the breaking of Lorentz invariance, since $V$ 
establishes the existence of a preferential reference frame for a background field by representing the vacuum energy related to the
cosmological constant within a de Sitter scenario\cite{Rodrigo}. 

Due to the wide implications of a background field connected to $V$, ranging from quantum mechanics\cite{N2012} to 
cosmological models\cite{N2016} with a potential of application in quantum gravity, important experiments attempting to either confirm 
or rule out this hypothesis would be very acknowledged. Then, here one proposes an accessible experiment in view of the current improvement of the technology 
of ultracold systems that has the goal of reaching temperatures in the order of $10^{-10}$K in CAL. 

\section{\label{sec:level1} A brief review of Symmetrical Special Relativity}

\begin{figure}
\includegraphics[scale=0.80]{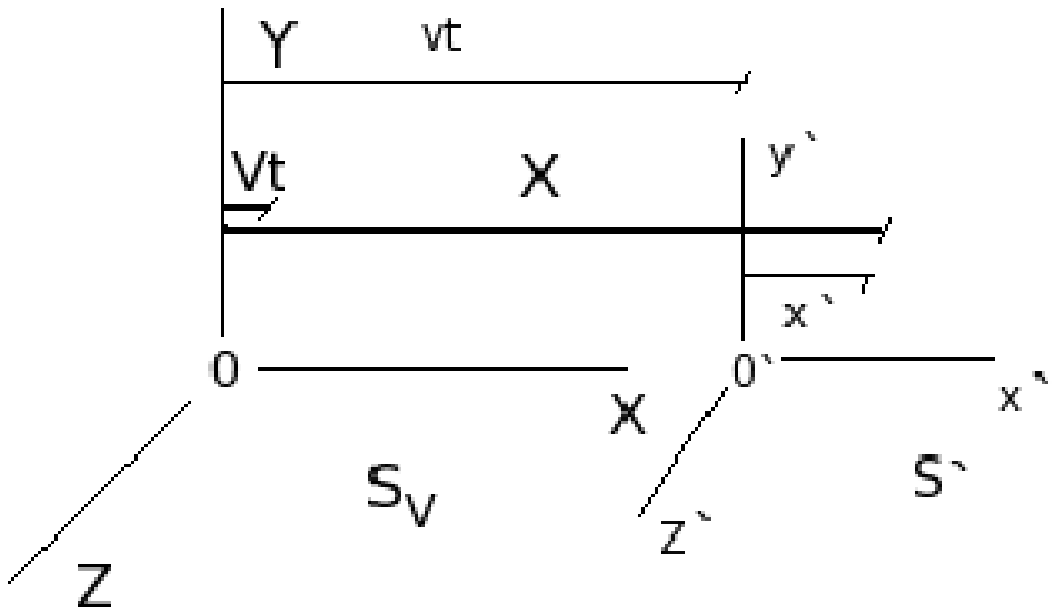}
\caption{In this special case $(1+1)D$, the referential $S^{\prime}$ moves in $x$-direction with a speed $v(>V)$ with respect to the
 background field connected to the ultra-referential $S_V$. If $V\rightarrow 0$, $S_V$ is eliminated (empty space), and thus the galilean
 frame $S$ takes place, recovering the Lorentz transformations.}
\end{figure}

Here some basic results of the Symmetrical Special Relativity (SSR)\cite{N2016} is presented. According to SSR, a new status of referential 
is defined, namely the so-called ultra-referential $S_V$, i.e., the background reference frame. In SSR, the idea of
ultra-referential is connected to the invariant minimum speed $V$, thus being unattainable by any particle with speed $v>V$.
Thus SSR provides the so-called {\it non-galilean reference frames}, since such a given frame can be thought of as being a set of all
the particles having a given speed $v$ with respect to the own ultra-referential $S_V$ of background field (Fig.1). 

The three postulates of SSR are the following:

1)~non-equivalence (asymmetry) of the non-Galilean reference frames due to the presence of the background frame $S_V$ that 
breaks down the Lorentz symmetry, i.e., we cannot change $v$ for $-v$ by means of inverse transformations in spacetime as we can never 
achieve the rest state for a certain non-Galilean reference frame in order to reverse the direction of its velocity just for one
spatial dimension. 

2)~the invariance of the speed of light ($c$).

3)~the covariance of the ultra-referential $S_V$ (the background framework) connected to an invariant minimum limit of speed ($V$).

 The $(1+1)D$-transformations in SSR\cite{N2016}\cite{N2015}\cite{N2012}\cite{N2010} with $\vec v=v_x=v$ (Fig.1) are 

\begin{equation}
x^{\prime}=\Psi(X-vt+Vt)=\theta\gamma(X-vt+Vt) 
\end{equation}

and 

\begin{equation}
t^{\prime}=\Psi\left(t-\frac{vX}{c^2}+\frac{VX}{c^2}\right)=\theta\gamma\left(t-\frac{vX}{c^2}+\frac{VX}{c^2}\right), 
\end{equation}
where the factor $\theta=\sqrt{1-V^2/v^2}$ and $\Psi=\theta\gamma=\sqrt{1-V^2/v^2}/\sqrt{1-v^2/c^2}$. 

The $(3+1)D$-transformations in SSR were shown in a previous paper\cite{N2016}, namely: 

\begin{equation}
\vec r^{\prime}=\theta\left[\vec r + (\gamma-1)\frac{(\vec r.\vec v)}{v^2}\vec v-\gamma\vec v(1-\alpha)t\right], 
\end{equation}
where $\alpha=V/v$. 

And 

\begin{equation}
t^{\prime}=\theta\gamma\left[t-\frac{\vec r.\vec v}{c^2}(1-\alpha)\right]. 
\end{equation}

From Eq.(3) and Eq.(4), we can verify that, if we consider $\vec v$ to be in the same direction of $\vec r$, with $r=X$, we recover 
the special case of $(1+1)D$-transformations given by Eq.(1) and Eq.(2). 

The breaking of Lorentz symmetry group in SSR destroys the properties of the transformations of Special Relativity (SR) and so generates
an intriguing deformed kinematics and dynamics for speeds very close to the minimum speed $V$, i.e., for $v\rightarrow V$. Thus, close
to $V$, relativistic effects to the reverse are found such as the contraction of the improper time and the dilation of space\cite{N2016}. In this new scenario,
the proper time also suffers relativistic effects such as its own dilation with respect to the improper time when 
$v\approx V$, thus leading to $\Delta\tau>\Delta t$. So it was obtained\cite{N2016}\cite{N2015}   

\begin{equation}
\Delta\tau\sqrt{1-\frac{V^{2}}{v^{2}}}=\Delta t\sqrt{1-\frac{v^{2}}{c^{2}}},
\end{equation}
from where it will be made an experimental prospect for detecting the new relativistic effect of {\it improper time contraction}
(or {\it proper time dilation}) close to the minimum speed $V(=\sqrt{Gm_em_p}e/\hbar\cong 4.58\times 10^{-14}m/s)$\cite{N2016} 
that breaks Lorentz symmetry at very low energies. We have $e=q_e/\sqrt{4\pi\epsilon_0}$, where $q_e$ is the electron charge. $m_e$ is 
the electron mass and $m_p$ is the proton mass.   

We should stress that the minimum speed $V$ was obtained in a previous paper\cite{N2016} by taking into account the well-known 
Dirac's large number hypothesis (LNH), where we have the ratio $F_e/F_g=q_e^2/4\pi\epsilon_0 G m_em_p\sim 10^{40}$. 

The structure of spacetime in SSR generates a new effect on mass (energy), being symmetrical to what happens close to the speed 
of light $c$, i.e., it was shown that $E=m_0c^2\Psi(v)=m_0c^2\sqrt{1-V^2/v^2}/\sqrt{1-v^2/c^2}$, so that $E\rightarrow 0$ when 
$v\rightarrow V$. It is noticed that $E=E_0=m_0c^2$ for $v=v_0=\sqrt{cV}$. The momentum is 
$p=m_0v\Psi(v)=m_0v\sqrt{1-V^2/v^2}/\sqrt{1-v^2/c^2}$ and the deformed dispersion relation is
$E^2=c^2p^2+m_0^2c^4(1-V^2/v^2)$\cite{N2016}\cite{N2015}. 

The spacetime metric of SSR is a deformed Minkowski metric with the scale factor $\Theta$\cite{N2016} working like 
a conformal metric\cite{Rodrigo}, as follows:  

\begin{equation}
 ds^{2}=\Theta\eta_{\mu\nu}dx^{\mu}dx^{\nu}, 
\end{equation}
where $\eta_{\mu\nu}$ is the Minkowski metric and $\Theta$ is a conformal factor. 

 In a recent paper\cite{Rodrigo}, it was shown that SSR-metric in Eq.(6) is equivalent to a conformal metric given in spherical coordinates 
 with a set of negative curvatures represented by a set of positive cosmological constants $\Lambda$ working like a cosmological scalar 
field within a de Sitter (dS) scenario, namely:

\begin{equation}
 ds^{2}= \frac{c^{2}dt^2}{\left(1-\frac{{\Lambda}r^2}{6c^2}\right)^2}-\frac{dr^{2}}{\left(1-\frac{{\Lambda}r^2}{6c^2}\right)^2}-
\frac{r^2d\Omega}{\left(1-\frac{{\Lambda}r^2}{6c^2}\right)^2}, 
\end{equation}
where $d\Omega=(d\theta)^2+\sin^2\theta(d\Phi)^2$ is the solid angle and the conformal factor is $\Theta=\theta^{-2}=1/(1-V^{2}/v^{2})
\equiv 1/(1-\Lambda r^2/6c^2)^2$\cite{Rodrigo} with coordinate dependence.   

\subsection{Energy barrier of the minimum speed connected to the vacuum energy: the idea of dressed mass}

\begin{figure}
\begin{center}
\includegraphics[scale=0.9]{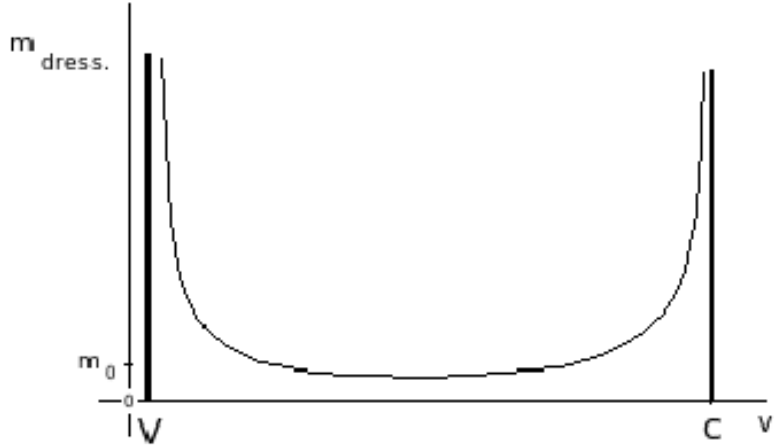}
\end{center}
\caption{The graph shows us two infinite barriers at $V$ and $c$, providing an aspect of symmetry of SSR. The first barrier ($V$) is exclusively due
 to the vacuum-$S_V$, being interpreted as a barrier of pure vacuum energy. In this regime we have the following approximations:
 $m_{eff}=m_{dressed}\approx\Delta m_{i}\approx m_0(1-V^2/v^2)^{-1/2}$ and  $m_r\approx m_0(1-V^2/v^2)^{1/2}$, so that
 $m_{dressed}\rightarrow\infty$ and $m=m_r=m_{bare}\rightarrow 0$ when $v\rightarrow V$. The second barrier ($c$) is a sum (mixture) of two contributions, namely the own
 bare
 (relativistic) mass $m$ that increases with the factor $\gamma=(1-v^2/c^2)^{-1/2}$ plus the interactive increment $\Delta
 m_{i}$ due to the
 vacuum energy-$S_V$, so that $m_{dressed}=m_L=m+\Delta m_{i}\approx m_0(1-v^2/c^2)^{-3/2}$. This is a longitudinal effect. For the
 transversal effect,
  $\Delta  m_{i}=0$ since we get $m_T=m$. This result will be shown elsewhere.}
\end{figure}

Let us consider a force applied to a particle, in the same direction of its motion. More general cases where the force is not 
necessarily parallel to velocity will be treated elsewhere. In our specific case ($\vec F||\vec v$), the relativistic power
$P_{ow}(=vdp/dt)$ of SSR is given as follows:

\begin{equation}
P_{ow}=v\frac{d}{dt}\left[m_0v\left(1-\frac{V^2}{v^2}\right)^{\frac{1}{2}}\left(1-\frac{v^2}{c^2}\right)^{-\frac{1}{2}}\right],
\end{equation}
where we have used the momentum in SSR, i.e., $p=m_0v\Psi(v)$.

After performing the calculations in Eq.(8), we find

\begin{equation}
 P_{ow}=\left[\frac{\left(1-\frac{V^2}{v^2}\right)^{\frac{1}{2}}}{\left(1-\frac{v^2}{c^2}\right)^{\frac{3}{2}}}
+\frac{V^2}{v^2\left(1-\frac{v^2}{c^2}\right)^{\frac{1}{2}}\left(1-\frac{V^2}{v^2}\right)^{\frac{1}{2}}}\right]
\frac{dE_k}{dt},
\end{equation}
where $E_k=\frac{1}{2}m_0v^2$.

If we make $V\rightarrow 0$ and $c\rightarrow\infty$ in Eq.(9), we simply recover the power obtained in newtonian mechanics, namely
$P_{ow}=dE_k/dt$. Now, if we just consider $V\rightarrow 0$ in Eq.(9), we recover the well-known relativistic power of SR, namely
$P_{ow}=(1-v^2/c^2)^{-3/2}dE_k/dt$. We notice that such a relativistic power tends to infinite ($P_{ow}\rightarrow\infty$) in the
limit $v\rightarrow c$. We explain this result as an effect of the drastic increase of an effective inertial mass close to $c$, namely
$m_{eff}=m_0(1-v^2/c^2)^{k^{\prime\prime}}$, where $k^{\prime\prime}=-3/2$. We must stress that such an effective inertial mass is the
response to an applied force parallel to the motion according to Newton second law, and it increases faster than the relativistic 
mass $m=m_r=m_0(1-v^2/c^2)^{-1/2}$.

  The effective inertial mass $m_ {eff}$ that we have obtained is a longitudinal mass $m_L$, i.e., it is a response to the force applied in the
 direction of motion. In SR, for the case where the force is perpendicular to velocity, we can show that the transversal mass increases like the relativistic
 mass, i.e., $m=m_T=m_0(1-v^2/c^2)^{-1/2}$, which differs from the longitudinal mass $m_L=m_0(1-v^2/c^2)^{-3/2}$.  So, in this sense, there is 
anisotropy of the effective inertial mass to be also investigated in more details by SSR in a further work.

 The mysterious discrepancy between the relativistic mass $m$ ($m_r$) and the longitudinal inertial mass $m_L$ from Newton second law
 (Eq.(9)) is a controversial issue\cite{Adler}\cite{Feynman}\cite{Okun}\cite{Sandin}\cite{Rindler}\cite{Taylor}. Actually the newtonian notion about inertia
 as the resistance to an acceleration
($m_L$) is not compatible with the relativistic dynamics ($m_r$) in the sense that we generally cannot consider $\vec F=m_{r}\vec a$. The 
 dynamics of SSR
 aims to give us a new interpretation for the inertia of the newtonian point of view in order to make it compatible with the relativistic
 mass. This compatibility is possible only due to the influence of the background field that couples to the particle and ``dresses" its relativistic mass in
 order to generate an
 effective (dressed) mass in accordance with the newtonian notion about inertia from Eq.(8) and Eq.(9). This issue will be
 clarified in this section.

 From Eq.(9), it is important to observe that, when we are closer to $V$, there emerges a completely new result (correction)
 for power, namely:

\begin{equation}
P_{ow}\approx\left(1-\frac{V^2}{v^2}\right)^{-\frac{1}{2}}\frac{d}{dt}\left(\frac{1}{2}m_0v^2\right),
\end{equation}
given in the approximation $v\approx V$. So, we notice that $P_{ow}\rightarrow\infty$ when $v\approx V$. We can also make the limit
$v\rightarrow V$ for the general case (Eq.(9)) and so we obtain an infinite power ($P_{ow}\rightarrow\infty$). Such a new relativistic effect deserves
the following very important comment:  Although we are in the limit of very low energies close to $V$, where the energy of the particle ($mc^2$) tends
to zero according to the approximation $E=mc^2\approx m_0c^2(1-V^2/v^2)^{k}$ with $k=1/2$, on the
other hand the power given in Eq.(10) shows us that there is an effective inertial mass that increases to infinite in the limit $v\rightarrow V$, that is to
say, from Eq.(10) we get the effective mass $m_{eff}\approx m_0(1-V^2/v^2)^{k^{\prime}}$, where $k^{\prime}=-1/2$. Therefore, from a dynamical point of view,
 the negative exponent $k^{\prime}$ ($=-1/2$) for the power at very low speeds (Eq.(10)) is responsible for the inferior barrier of the
 minimum speed $V$, as well as the exponent $k^{\prime\prime}=-3/2$ of the well-known relativistic power is responsible for the top barrier
 of the speed of light $c$ according to Newton second law (Fig.2). Actually, due to the drastic increase of $m_{eff}$ of a particle moving
 closer to $S_V$, leading to its strong coupling to the vacuum field in the background frame $S_V$, thus, in view of this, the dynamics 
 of SSR states that it is impossible to decelerate a subatomic particle until reaching the rest. This is the reason why there is a
 unattainable minimum speed $V$. 

 In order to see clearly both exponents $k^{\prime}=-1/2$ (inferior inertial barrier $V$) and $k^{\prime\prime}=-3/2$ (top inertial barrier
 $c$), let us write the general formula of power (Eq.(9)) in the following alternative way after some algebraic manipulations on it,
 namely:

\begin{equation}
P_{ow}=\left(1-\frac{V^2}{v^2}\right)^{k^{\prime}}\left(1-\frac{v^2}{c^2}\right)^{k^{\prime\prime}}\left(1-\frac{V^2}{c^2}\right)
\frac{dE_k}{dt},
\end{equation}
where $k^{\prime}=-1/2$ and $k^{\prime\prime}=-3/2$. Now it is easy to see that, if $v\approx V$ or even $v<<c$, Eq.(11) recovers the
approximation in Eq.(10). As $V<<c$, the ratio $V^2/c^2(<<1)$ in Eq.(11) is a very small dimensionless constant 
$\xi^2=V^2/c^2\sim 10^{-44}$\cite{N2016}. So $\xi^2$ can be neglected in Eq.(11).

So, from Eq.(11) we get the effective inertial mass $m_{eff}$ of SSR, namely:

\begin{equation}
m_{eff}=m_0\left(1-\frac{V^2}{v^2}\right)^{-\frac{1}{2}}\left(1-\frac{v^2}{c^2}\right)^{-\frac{3}{2}}. 
\end{equation}

We must stress that $m_{eff}$ in Eq.(12) is a longitudinal mass $m_L$. The problem of mass anisotropy will be treated elsewhere. But here
we will intend to show that, just for the approximation $v\approx V$, the effective inertial mass becomes practically isotropic, that is to
say $m_L\approx m_T\approx m_0\left(1-\frac{V^2}{v^2}\right)^{-1/2}$. This important result will show us the isotropic aspect of the
 vacuum-$S_V$, 
so that the inferior barrier $V$ has the same behavior of response ($k^{\prime}=-1/2$) of a force applied at any direction
in the space, namely for any angle between the applied force and the velocity of the particle.

 We must point out the fact that $m_{eff}$ has nothing to do with the ``relativistic mass" (relativistic energy $E=m_0c^2\Psi(v)$) in the
 sense that
$m_{eff}$ is dynamically responsible for both barriers $V$ and $c$. The discrepancy between the ``relativistic mass" (energy $mc^2$ of the particle) and
such an effective inertial mass ($m_{eff}$) can be interpreted under SSR theory, as follows: $m_{eff}$ is a dressed inertial mass given 
in response to the presence of the vacuum-$S_V$ that works like a kind of ``fluid" in which the particle $m_0$ is immersed, 
while the ``relativistic mass" in SSR ($m=m_{rel}=m_0\Psi(v)$) works like a
 bare inertial mass in the sense that it is not considered to be under the dynamical influence of the ``fluid" connected to the
 vacuum-$S_V$. That is the reason
 why the exponent $k=1/2$ cannot be used to explain the existence of an infinite barrier at $V$, namely the vacuum-$S_V$
 barrier is governed by
the exponent $k^{\prime}=-1/2$ as shown in Eq.(10), Eq.(11) and Eq.(12), which prevents $v_*(=v-V)\leq 0$.

 The difference betweeen the dressed (effective) mass and the relativistic (bare) mass, i.e., $m_{eff}-m$ represents an interactive
 increment of mass $\Delta m_{i}$ that has purely origin from the vacuum energy of $S_V$, mamely: 

\begin{equation}
\Delta m_{i}= m_0\left[\frac{\left(1-\frac{V^2}{c^2}\right)}{\left(1-\frac{V^2}{v^2}\right)^{\frac{1}{2}}
\left(1-\frac{v^2}{c^2}\right)^{\frac{3}{2}}}-\frac{\left(1-\frac{V^2}{v^2}\right)^{\frac{1}{2}}}{\left(1-\frac{v^2}{c^2}\right)^{\frac{1}{2}}}\right]
\end{equation}

We have $\Delta m_{i}=m_{eff}-m$, where $m_{eff}=m_{dressed}$ is given in Eq.(12) and $m$ ($m_r=m_0\Psi$) is the relativistic mass
in SSR.

 From Eq.(13), we consider the following special cases:

a) for $v\approx c$ we have

\begin{equation}
 \Delta m_{i}\approx m_0\left[\left(1-\frac{v^2}{c^2}\right)^{-\frac{3}{2}}-\left(1-\frac{v^2}{c^2}\right)^{-\frac{1}{2}}\right]
\end{equation}

As the effective inertial mass $m_{eff}$ ($m_L$) increases much faster than the bare (relativistic) mass $m$ ($m_r$) close to the speed $c$,
 there is an increment of inertial mass $\Delta m_i$ that dresses $m$ in direction of its motion and it tends to be infinite when $v\rightarrow c$,
i.e., $\Delta m_i\rightarrow\infty$.

b) for $V<<v<<c$ (newtonian or intermediary regime) we find $\Delta m_i\approx 0$, where we simply have $m_{eff}$ ($m_{dressed}$)$\approx m\approx m_0$.
 This is the classical approximation.

c) for $v\approx V$ (close to the vacuum-$S_V$ regime), we have the following approximation:

\begin{equation}
\Delta m_{i}=(m_{dressed}-m)\approx m_{dressed}\approx\frac{m_0}{\sqrt{1-\frac{V^2}{v^2}}},
\end{equation}
where $m=m_0\Psi\approx 0$ when $v\approx V$.

The approximation in Eq.(15) shows that the whole dressed mass has purely origin from the energy of vacuum-$S_V$, with $m_{dressed}$ being
the pure increment $\Delta m_{i}$, since the bare (relativistic) mass $m$ of the own particle
 almost vanishes in such a regime ($v\approx V$), and thus an inertial effect only due to the vacuum (``fluid")-$S_V$
remains. We see that $\Delta m_{i}\rightarrow\infty$ when $v\rightarrow V$. In other words, we can interpret this infinite barrier of vacuum-$S_V$ by
considering the particle to be strongly coupled to the background field-$S_V$ in all directions of the space. The isotropy of $m_{eff}$ in this regime will
be shown in detail elsewhere, being $m_{eff}=m_L=m_T\approx m_0(1-V^2/v^2)^{-1/2}$. In such a regime, the particle practically loses its locality
(``identity") in the sense that it is spread out isotropically in the whole space and it becomes strongly coupled to the vacuum field-$S_V$, leading to an
 infinite value of $\Delta m_{i}$. Such a divergence of the dressed mass has origin from the dilation factor $\Theta_v(\rightarrow\infty)$ for this regime
when $v\approx V$, so that we can rewrite Eq.(15) in the following way: $\Delta m_{i}\approx m_{dressed}\approx m_0\Theta(v)^{1/2}$. That is
essentially the dynamical explanation why the particle cannot reach the rest in SSR theory so that the background frame of 
the vacuum-$S_V$ becomes unattainable for any particle. However, in the macroscopic (classical) level, the
minimum speed $V$ as well as the Planck constant $\hbar$ are neglectable as a good approximation, such that the rest state 
is naturally recovered in spite of the subatomic particles that constitute a macroscopic object at rest are always moving, since its 
temperature can never reach the absolute zero\cite{N2010}, nor their constituent subatomic particles can ever reach $V$.

\section{\label{sec:level1} The proper time dilation of an atomic clock in a ultracold gas}

Let us consider a ultracold gas so that the root mean square speed per atom (molecule) of the gas is close to the minimum speed
$V$, i.e., in this special case, all the ``atoms'' (`` molecules'') are found in the same state with speed $v$ close to $V$, so that one
can write $\sqrt{\left<v^2\right>}=v\rightarrow V$. Here it must be stressed that when one is too close to such a universal 
minimum speed, the idea of an individual atom (molecule) with mass $m_0$ in a gas does not make sense, as one is very close to the
vacuum regime working like a kind of superfluid (condensate) represented by a ``super-atom'' that plays the role of a
super-particle of vacuum with Planck mass ($M_P$), i.e., a kind of dark energy could be produced in such a ultra-special condition 
of too low temperatures as expected to occur in CAL. And SSR comes to predict a change in passing of the time for this regime of Lorentz
violation, i.e., we find Eq.(5) given in the approximation $v\approx V$ ($\Delta\tau\approx\Delta t\theta^{-1}=\Delta t(1-V^2/v^2)^{-1/2}$). 

In order to justify the Planck mass ($M_P$) for representing the mass of the ``super-atom'' (super-particle) as being the set of all the
strongly correlated particles given in a regime too close to the minimum speed $V$, let us think that the strong correlation between the
particles so close to $V$ can be explained by the increasing of the dressed mass of any particle of the gas, so that we consider the 
existence of a upper cut-off for the dressed mass, being associated with $M_P\sim 10^{-8}Kg$, i.e., $M_P\sim m_{dressed}=m_0\theta^{-1}(v\approx V)$ 
(Eq.(15)), since we should have in mind that the Planck energy $E_P=M_Pc^2\sim 10^{19}$ GeV is the highest scale of energy within the
cosmological scenario. So, we should note that $M_P$ is the standard universal mass connected to vacuum. 

The dilation of the proper time increases the persistence of correlation between two points inside the condensate and thus keeps the 
synchronization of two atomic clocks even out of the thermal equilibrium where a very rapid energy dissipation occurs between both 
points (atomic clocks) as shown recently by Pigneur et al\cite{Marine} who claim that such a persistence of correlation out of the
equilibrium is mantained by a mechanism still unknown and whose origin could be elucidated by means of the proper time dilation of both 
atomic clocks inside the condensate. Thus the spacetime of SSR seems to be the foundation to explain the experimental evidence reported in
the recent reference\cite{Marine}, but this subject deserves a more careful investigation elsewhere. 

Later it will be realized that there are two extreme energy regimes where matter is condensed into two types of vacuum, one of 
them with very high temperature (Planck energy $M_Pc^2$ for the minimum length of Planck $L_P$) and the other one with a very low 
temperature to be investigated as consequence of SSR, where such a lowest speed $V$ would be related to a very low temperature close to the 
absolute zero temperature, which is considered to be in accordance with the classical kinematics like newtonian or even relativistic
mechanics, where there exist rest connected to the absolute zero temperature. Such inconsistency between classical motion (rest) and
the absolute zero temperature, which is prevented by the third law of thermodynamics can be solved by SSR, since the non-classical 
mechanics of SSR with a unattainable minimum speed $V$ could justify dynamically the impossibility of reaching 
the absolute zero temperature\cite{N2010}. But, on the other hand, one should be pointed out to the fact that the absolute zero 
temperature would mean a complete absence of motion as well as an infinite temperature would be related to an infinite speed, which must
be replaced by the finite speed of light $c$. In this sense,
the absolute zero temperature as the absence of motion does not seem to be compatible with the existence of a non-null minimum 
speed $V$ according to SSR, which is able to introduce a compatibility between the lowest speed $V$ and a lowest
non-null temperature ($T_{min}$) by correcting by a lower cut-off the absolute zero as well as it is already known that the speed of light
$c$ comes to replace an infinite speed, and an infinite temperature by a upper cut-off of maximum temperature so-called
Planck temperature $T_P(\sim 10^{32}K)$. 

Therefore SSR also introduces two cut-offs of temperature and energy in the cosmological scenario, namely 
one of them is well-konwn as the Planck energy in the early universe related to the Planck length $L_P$, i.e., a UV cut-off ($E_P=M_Pc^2$),  
and the other one would be related to a minimum energy density of vacuum connected to a horizon (maximum) length $L_{\Lambda}$ for a given 
cosmological constant ($\Lambda$), i.e., a IR cut-off given by SSR ($E_{min}=M_PV^2$), as it was initially investigated by
Padmanabhan\cite{Pad} and recently advanced by SSR that has introduced an invariant minimum speed in order to provide a
kinematics explanation for the cosmological constant\cite{N2016}\cite{N2015} and also for dS-horizon $L_{\Lambda}$\cite{Rodrigo} 
connected to $E_{min}$ for representing the most cold cosmological vacuum. In order to perceive this as a classical picture, imagine a 
super-particle of vacuum with mass $M_P$ having a minimum vibration $V$. On the other hand, $E_{P}(\sim 10^{19}GeV$) represents the most hot cosmological vacuum 
(early universe) with the strongest repulsive field well-known as {\it inflaton}, but this issue should be deeply treated by using the
formalism of energy-momentum tensor in SSR elsewhere. 

The theory of SSR establishes a symmetry of both speeds and temperatures, so that we have $V<v<c$ for representing $T_{min}<T<T_{P}$, where such
a mininum temperature $T_{min}$ should be determined by SSR, since there should be a connection between a unattainable non-zero
temperature $T_{min}$ and the new universal constant of minimum speed $V$. Now it is easy to realize that such connection is obtained
analogously to that one already obtained for the maximum temperature $T_P$ and $c$, where $T_P=M_Pc^2/K_B (\sim 10^{32}K)$, with 
$M_P(=\sqrt{\hbar c/G}\cong 2.176\times 10^{-8}kg)$ being the Planck mass and $K_B(\cong 1.38\times 10^{-23}J/K)$ the Boltzmann constant.  
As it is known that $T_P\propto c^2$, due to the symmetry of both limits of speed in SSR, it is expected that $T_{min}\propto V^2$,
so that it is found that $T_{min}=M_PV^2/K_B (\cong 3.28\times 10^{-12}K)$ and thus one obtains the ratio
$T_{min}/T_{P}=V^2/c^2\sim 10^{-44}$. In view of this minimum ratio and by considering a gas with temperature $T$, so the two following
fundamental equivalence relations are obtained as:    

\begin{equation}
\frac{V^2}{\left<v^2\right>}=\frac{T_{min}}{T},
\end{equation}
so that $T>T_{min}$ for $\sqrt{\left<v^2\right>}>V$. 

Here it must be stressed that $T_{min}$ comes to replace the classical idea of an absolute zero by an absolute non-null vibration due
to the invariant minimum speed $V$, so that now one cannot find $T\leq T_{min}$ or even $v\leq V$ within the cosmological scenario
with temperature. Thus SSR allows to define a cosmological (absolute) scale of temperature where the lowest temperature is of the
order of $10^{-12}$K. Let us call it as the {\it Planck scale of temperature} in SSR.  

On the other hand, it is easy to obtain

\begin{equation}
\frac{\left<v^2\right>}{c^2}=\frac{T}{T_P},
\end{equation}
where we have $T<T_{P}$ for $\sqrt{\left<v^2\right>}<c$. 

Here it must be also stressed that one considers the cosmological scale of temperature according to the Standard Model (SM) since the
maximum temperature is the Planck temperature ($T_P$) that unifies all the $4$ forces of nature. On the other hand, the minimal temperature 
($T_{min}\propto V^2$) that has origin in a quantum gravity at lower energies close to $V$ is beyond SM of cosmology due to 
its direct connection with the dark energy and also the inflationary field (inflaton), but this subject should be explored elsewhere.   

As only the new effect of proper time dilation close to the minimum speed $V$ or its equivalent minimum temperature
$T_{min}(=M_PV^2/K_B\sim 10^{-12}K)$ could be tested in CAL (a Lorentz violation in the infrared regime), we make the following 
approximation of vacuum ($m_{dressed}\approx M_P$) given for gases with much lower temperatures in Eq.(5), i.e., we just consider 
$T\approx T_{min}$ ($\sqrt{\left<v^2\right>}\approx V$), so that we just take into account Eq.(16) into Eq.(5) and thus we neglect Eq.(17) 
($T/T_P\approx 0$), by simply writing Eq.(5) in the form, as follows: 

\begin{equation}
\tau\approx\frac{t}{\sqrt{1-\frac{V^2}{\left<v^2\right>}}}=\frac{t}{\sqrt{1-\frac{T_{min}}{T}}}, 
\end{equation}
where $T_{min}=M_PV^2/K_B$, with $\theta\equiv\mathcal\theta(T)=\sqrt{1-\frac{T_{min}}{T}}$. 

We have $\tau$, which is the proper time of an atomic clock (atom of a radioactive sample) inside the ultracold gas (condensate), 
and $t$ is the improper time given in lab at room temperature. 

\subsection{Suppression of the half-life time}

\begin{figure}
\includegraphics[scale=1.0]{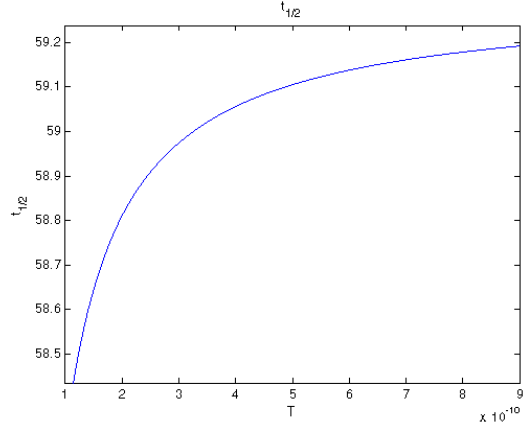}
\caption{The graph shows the suppression of the half-life time [Eq.(22)] for $Na^{25}$ thermalized with a 
ultracold gas of $Na^{23}K^{40}$ for the range of temperatures $10^{-10}K\leq T\leq 10^{-9}K$ close to the order of temperature that 
will be obtained in CAL. It is known that $t_{1/2}\cong 59.3$s for the radioactive sample of $Na^{25}$ with $\beta^{-}$ decay, where 
$Na^{25}\rightarrow\beta^{-}+Mg^{25}$.}
\end{figure}

\begin{figure}
\includegraphics[scale=1.0]{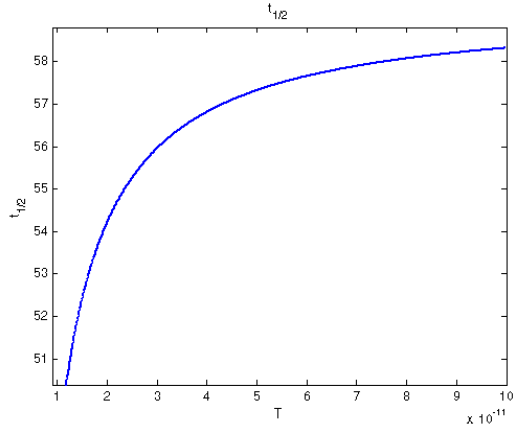}
\caption{The graph shows the suppression of the half-life time [Eq.(22)] for $Na^{25}$ thermalized with a 
ultracold gas of $Na^{23}K^{40}$ for the range of temperatures with one order below of that one obtained by the current technology, i.e., 
$10^{-11}K\leq T\leq 10^{-10}K$, which will be probably obtained in CAL after a technological improvement.}
\end{figure}

By considering initially $N_0$ atoms of a radioactive sample that begins to decay, then, after the time $t$ in lab, such a sample will 
acquire a number $N$ of atoms, so that 

\begin{equation}
 N=N_0\exp(-\lambda t), 
\end{equation}
where $\lambda$ is the so-called disintegration constant of a given radioactive sample. The time $t_{1/2}$ that leads to $N=N_0/2$ is
well-known as the half-life time and it is given by $t_{1/2}=ln(2)/\lambda$. Normally the half-life time depends only on the type of 
radioactive sample given by a certain disintegration constant $\lambda$, since one considers that the proper time $\tau$ of an atomic
clock inside a ultracold sample is equal to the improper one $t$ in lab. However it will be shown that $t_{1/2}$ is suppressed by lower 
temperatures due to the dilation of the proper time ($\tau>t$) according to Eq.(18) that reveals Lorentz violation in a ultra-cold gas,
thus leading to an effective disintegration constant with temperature dependence. To see the temperature effect on half-life time, 
Eq.(19) is written as   

\begin{equation}
 N=N_0\exp(-\lambda_Tt)=N_0\exp\left(-\frac{\lambda t}{\sqrt{1-\frac{T_{min}}{T}}}\right), 
\end{equation}  
where there emerges an effective disintegration constant $\lambda_T$ due to the dilation of the proper time [Eq.(18)] close to the
minimum temperature $T_{min}$ connected to the minimum speed $V$, so that we find $\lambda_T=\lambda/\sqrt{1-T_{min}/T}$, which means
an increase of the desintegration constant at lower temperatures, thus leading to the half-life time suppressed by very low temperatures,
as:

\begin{equation}
 t_{1/2}(T)=\frac{ln(2)}{\lambda_T}=\frac{ln(2)}{\lambda}\sqrt{1-\frac{T_{min}}{T}}, 
\end{equation} 

or simply 

\begin{equation}
 t_{1/2}^{\prime}=t_{1/2}\sqrt{1-\frac{T_{min}}{T}}, 
\end{equation} 
where $t_{1/2}(T)=t_{1/2}^{\prime}$ is the suppressed half-life time of the ultracold radioactive sample (e.g: $Na^{25}$) and $t_{1/2}$
is the well-known half-life time. 

The suppressed half-life given in Eq.(22) and shown in Fig.3 and Fig.4 for two intervals of very low temperatures closer to $10^{-12}$K
represents the experimental signature of a Lorentz violation in the infrared regime due to the universal minimum speed $V$.

The equation for the decay law with a half-life time suppression is 

\begin{equation} 
\frac{dN}{dt}=-R(T)=-\frac{\lambda}{\sqrt{1-\frac{T_{min}}{T}}}N,
\end{equation}
whose solution is given by Eq.(20), thus leading to a suppression of the half-life time at lower temperatures [Eq.(22)] 
as it can be seen in Fig.3 and Fig.4 for $\beta^{-}$ decay of the isotope $Na^{25}$. $R(T)=-dN/dt$ is the increased decay rate for
lower temperatures. So we write 

\begin{equation} 
R(T)=\frac{\lambda}{\sqrt{1-\frac{T_{min}}{T}}}N. 
\end{equation}

In Eq.(22), when $T\rightarrow T_{min}(\sim 10^{-12}K)$, the half-life time of the radioactive sample would become completely 
suppressed ($t_{1/2}^{\prime}\rightarrow 0$), but the best technology of cooling will be made soon by NASA in CAL for reaching the order of $10^{-9}K$
of a sodium-potassium gas confined during a few of seconds. This first experiment could lead to a very slight suppression of the half-life 
time of a certain radioactive sample thermalized with the sodium-potassium gas ($Na^{23}K^{40}$) and thus the detection of such a suppression 
seems to be difficult at a first sight depending on the time resolution of the cooling devices with {\it lasers}. However, in the future,
the improvement of the experiment in CAL for reducing even more the temperature until reaching the order of 
$10^{-10}K$ with confination of a few hundred of seconds will allow to detect a larger suppression of the half-life time as, for instance, 
a sample of the isotope $Na^{25}$ whose half-life time is about $t_{1/2}\cong 59.3$s with the usual desintegration constant 
$\lambda\cong 0,0117s^{-1}$, thus needing a time of confination of a few hundred of seconds to be at least of the same order of its 
half-life time.
   
\section{Prospects}

The study of the ultracold atoms close to the minimum speed $V$, which has origin in gravity ($V\propto\sqrt{G}$)\cite{N2016} could 
reshape our understanding of matter, dark energy and the fundamental nature of gravity within the scenario of a quantum gravity
connected to the cosmological constant. So the experiments that will be made in CAL could give us insight into a new aspect of quantum 
gravity and dark energy by means of a change in the passing of time predicted by SSR at much lower energies by breaking down Lorentz symmetry,
thus leading to the half-life time suppression of a ultracold radioactive sample (e.g: $Na^{25}$) thermalized with a ultracold gas 
(e.g: $Na^{23}K^{40}$). Other radioactive samples or even ultracold gases could be used depending on the most favorable experimental 
conditions.

\end{document}